\makeatletter
\declare@file@substitution{revtex4-1.cls}{revtex4-2.cls}
\makeatother
\documentclass{aastex631}
\usepackage{amsmath}
\usepackage{graphicx}
\begin{document}


\title{Multi-wavelength Emission of Gamma-ray Burst Prompt Phase. \\II. Spectral Polarimetry}


\author{Jia-Sheng Li}
\author[0000-0001-5641-2598]{Mi-Xiang Lan}
\affiliation{Center for Theoretical Physics and College of Physics, Jilin University, Changchun, 130012, China; lanmixiang@jlu.edu.cn}

\begin{abstract}
Polarization spectra had been predicted within the photosphere model. For the purpose of seeking more clues to distinguish between the models, both the time-resolved and time-integrated polarization spectra from optical band to MeV gamma-rays of the magnetic reconnection model are studied here. There are two newly found differences between the two models. First, the time-integrated polarization degree (PD) of the magnetic reconnection model would in general increase with frequency for on-axis observations, while it is not monotonous for the photosphere model. Second, the variations of both the time-integrated and the time-resolved polarization angles (PAs) with frequency of the magnetic reconnection model is not random, while the time-integrated PA varies randomly with frequency for the photosphere model. Therefore, future energy-resolved polarization analysis could distinguish between the two models. In addition, the PA rotation spectra are studied for the first time. The rotation value of PA within the burst duration will decrease with the increase of the observational energy band. Most significant PA rotation would happen for slightly off-axis observations in each energy band. The PA would rotate even for on-axis observations in optical band. Compared with the aligned magnetic field case, the PA rotation is quite rare in the gamma-ray band for the case with a toroidal field in the radiation region.

\end{abstract}


\section{Introduction}

Gamma-ray bursts (GRBs) are the high-energy radiation produced by the collapse of a massive star \citep{1993ApJ...405..273W,1999Natur.401..453B,2001ApJ...550..410M,2003ApJ...599L..95M,2003Natur.423..847H} or the merger of the two compact stars \citep{1992ApJ...395L..83N,2017ApJ...848L..13A,2017ApJ...848L..14G,2018PhRvL.120x1103L}. At present, there are three main models for the GRBs prompt phase: the internal shock model \citep{1992ApJ...395L..83N, PX1994, 1994ApJ...430L..93R, Daigne2009}, the photospheric model \citep{1994MNRAS.270..480T,2000ApJ...529..146E,2000ApJ...530..292M,2005ApJ...628..847R,2009ApJ...700L..47L,2011ApJ...737...68B,2011ApJ...732...49P,2011ApJ...732...26M,2011ApJ...731...80N,2012ApJ...746...49X,2013ApJ...772...11R,2013ApJ...767..139B,2013MNRAS.428.2430L,2013ApJ...765..103L}, and the magnetic reconnection model \citep{2008A&A...480..305G,2011ApJ...726...90Z,2016MNRAS.459.3635B,2016ApJ...816L..20G}.

Since both the light curves and energy spectra in GRB prompt phase could be interpreted by these three models, these observable quantities could not be used to distinguish the models. The polarization provides two additional observable quantities, the polarization degree (PD) and the polarization angle (PA), which are sensitive to radiation mechanisms \citep{Toma_2009, Lundman_2018, Lan_2020, Parsotan_2020, Gill2020, Lan_2021b, Gill_2021, Parsotan_2022, Guan_2023, SL2024, 2024A&A...687A.128C}. It is predicted that the PD value in the gamma-ray band is roughly zero for the photosphere model \citep{Lundman_2018, Parsotan_2020, Parsotan_2022}. However, an upper limit of $\sim(40-50)\%$ in the magnetic reconnection model is predicted \citep{SL2024, Li_2024}. In addition, PA evolves randomly in both the optical and gamma-ray bands for the photosphere model \citep{Parsotan_2022}, while it is not random in the gamma-ray band for the synchrotron model \citep{Lan_2020, Lan_2021b, WL_2023a, WL_2023b, Li_2024}. Therefore, up till now we have two method to distinguish between the models in GRB prompt phase: one is the PD value at gamma-ray band and the other is the PA evolution pattern. 

In addition to the polarization curves, polarization spectra in GRB prompt phase were also studied \citep{Lan_2020,  Lan_2021b, Parsotan_2022}. The time-resolved PD will increase with frequency from X-rays to MeV gamma-rays in the magnetic reconnection model \citep{Lan_2020}. In the photosphere model, the time-integrated PD will decrease with frequency in optical band, then there is a PD peak around $10^{-1}$ keV, in higher energy band (around $[10-10^2]$ keV) the time-integrated PD is zero and finally it will increase with frequency beyond $(10^2-10^3)$ keV \citep{Parsotan_2022}. In order to provide more clues to distinguish between the two models, the time-integrated PD spectra should be studied for the magnetic reconnection model.

Under frame work of the magnetic reconnection model, the predicted time-resolved PA is a constant with frequency from X-rays to 10 MeV gamma-rays for on-axis observation of the axisymmetric jet \citep{Lan_2020}. In high-energy gamma-ray band, the rotation of the PA with time is most likely to happen for slightly off-axis observations \citep{WL_2023a} and then three key parameters that have significant influence on the PA rotations were found \citep{WL_2023b}. And the maximum rotation value of the time-resolved PA within the burst duration is $90^\circ$ in gamma-ray band \citep{WL_2023b}. For the photosphere model, the predicted time-integrated PA vary randomly with frequency at both optical band and gamma-ray band \citep{Parsotan_2022}. However, under frame work of the magnetic reconnection model, the laws for PA rotation in the lower energy band during GRB prompt phase is not yet known and whether or not the change of the time-resolved PA within the burst duration vary with frequency is also unknown. In addition, how the time-integrated PA vary with frequency from the optical band to gamma-rays has also not been invesigated so far.

Recently, the polarization spectra were studied using the POLAR's data \citep{2023arXiv230900507D}. Due to the low statistics and the low energy resolution, they did not found significant dependence of the PD and PA on the energy. In recent years, polarization detectors in various enengy bands are now in comission or will be launched soon. In $\gamma$-ray band, both the High-energy Polarimetry Detector (HPD) \citep{POLAR2} on board POLAR-2 and the Compton Spectrometer and Imager (COSI) \citep{2019BAAS...51g..98T} are planned to launch in the near future. In the X-ray band, the Imaging X-ray Polarimetry Explorer (IXPE) \citep{Negro2023} is now in comission and the Low-energy Polarimetry Detector (LPD) \citep{POLAR2} on board POLAR-2 is planned. The Very Large Telescope (VLT) and Liverpool Telescope (LT) are the polarization telescopes working in optical band. Therefore, there will be abundant multi-wavelength data in the near future. It is necessary to predict the multi-wavelength polarization (including the polarization curves \footnote{Polarization curves refer to the variation of the energy-resolved or energy-integrated polarization with time, including the PD curve and PA curve.} and polarization spectra \footnote{Polarization sepctra refer to the variation of the time-resolved or time-integrated polarization with energy, including the PD spectrum and PA spectrum.}) to interpret these multi-wavelength polarization data and then to distinguish between the models.

In a previous paper, we studied the multi-wavelength polarization curves and the influence of the key parameters on the time-integrated polarizations \citep{Li_2024}. Here, the polarization spectra, including the PD spectra, the PA spectra and the PA rotation spectra, are studied. The model is presented in Section \ref{sec:The Models}. In Section \ref{sec:Numerical Results}, we show time-resolved and time-integrated polarization spectra in the GRB prompt phase, and the influence of key parameters on polarization spectra are also investigated. Finally, we present conclusions and discussion in Section \ref{sec:Conclusions and Discussion}.

\section{The Models}\label{sec:The Models}

As in the previous studies \citep{2015ApJ...808...33U,2016ApJ...825...97U,2018ApJ...869..100U}, the radiation region is assumed to be a relativistic thin shell expanding radially from the central engine at redshift $z$. The electrons in the shell are accelerated via the magnetic reconnection process to emit synchrotron  photons in the magnetic field. The shell is assumed to begin the radiation at radius $r_{on}$ and stop at $r_{off}$. And the shell is also accelerated and the variation of its bulk Lorentz factor $\Gamma$ is a power-law with the radius $r$ from the central engine \citep{2002A&A...387..714D}.
\begin{equation}
\Gamma(r)=\Gamma_0(r/r_0)^s,
\end{equation}
where we take $s=0.35$ for an aligned field in the shell, but for a toroidal field the shell roughly maintains a constant bulk Lorentz factor and $s=0$ \citep{2002A&A...387..714D}. The $\Gamma_0$ is the normalization value of the bulk Lorentz factor at the reference radius $r_0$. The parameters marked with “ $'$ ” in this paper are the comoving-frame quantities. And correspondingly, the magnetic field strength $B'$ of the shell decays with the radius $r$ and can be expressed as:
\begin{equation}
B'(r)=B'_0(r/r_0)^{-b},
\end{equation}
where the decay index $b$ is taken 1 \citep{2002A&A...387..714D}. The $B'_0$ is the normalization value of the magnetic field strength at the reference radius $r_0$.

We assume the shape of the photon spectrum $H_{en}(\nu')$ consists of three-segment power-laws, and the expression is as follows:
\begin{equation}\label{Hen}
H_{en}(\nu')=\begin{cases}
(\nu'/\nu'_1)^{\alpha_1+1}, & \text{$\nu'< \nu'_1$}, \\ (\nu'/\nu'_1)^{\alpha_2+1}, & \text{$\nu'_1< \nu'< \nu'_2$}, \\ (\nu'_2/\nu'_1)^{\alpha_2+1}(\nu'/\nu'_2)^{\beta+1}, & \text{$\nu'> \nu'_2$},
\end{cases}
\end{equation}
where $\nu'$, $\nu'_1$ and $\nu'_2$ equal to $\nu(1+z)/\mathcal{D}$, $\min(\nu'_{cool}, \nu'_{min})$ and $\max(\nu'_{cool}, \nu'_{min})$, respectively. The $\nu$ is the observational frequency and $\mathcal{D}=1/\Gamma/(1-\beta_v\cos\theta)$ is the Doppler factor, where $\beta_v$ is the corresponding dimensionless velocity and $\theta$ is the angle between the line of sight and the local velocity direction. 

The $\alpha_1$, $\alpha_2$ and $\beta$ are the low-energy, mid-energy and high-energy photon spectral indices, respectively. If the GRB is in a slow-cooling phase (i.e. $\nu'_{min}< \nu'_{cool}$), $\alpha_1=-2/3$, $\alpha_2=-1-\dfrac{p-1}{2}$ and $\beta=-1-\dfrac{p}{2}$. The $p$ is the index of the true energy spectrum ($N(\gamma_e)\propto\gamma^{-p}_e$) of the injected electrons in the shell, and its value is taken as $2.6$ here. If the GRB is in a fast-cooling phase (i.e. $\nu'_{cool}< \nu'_{min}$), $\alpha_1=-2/3$ and $\beta=-1-\dfrac{p}{2}$. The value of $\alpha_2$ in the fast-cooling phase tends to $-3/2$ in a stronger decaying magnetic field and to $-1$ when the magnetic field is weaker \citep{2014NatPh..10..351U}. Observationally, the value of $\alpha_2$ is concentrated around $-1$ in the Band-function \citep{2016A&A...588A.135Y,2021ApJ...913...60P}, where the fitting is done above the energy band of the soft X-rays. However, it is around $-3/2$ when the analysis energy band is extended down to the optical band  \citep{2017ApJ...846..137O,2018A&A...616A.138O,2019A&A...628A..59O,2019A&A...625A..60R,2021A&A...652A.123T}. So the two values of the mid-energy photon spectral index $\alpha_2$ are considered in fast-cooling case. 

The $\nu'_{cool}$ and $\nu'_{min}$ read:
\begin{equation}
\nu'_{cool}=\frac{q_eB'\gamma_{cool}^2\sin\theta'_B}{2\pi m_ec},\ \ \ \ \nu'_{min}=\frac{q_eB'\gamma_{ch}^2\sin\theta'_B}{2\pi m_ec},
\end{equation}
where $m_e$ and $q_e$ are the mass and charge of the electron, respectively. $c$ is the speed of light. The $\theta'_B$ is the pitch angle of the electrons. $\gamma_{cool}=(6\pi m_ec\Gamma)/(\sigma_TB'^2t)$ is the cooling Lorentz factor of electrons and $t$ is the dynamical time in the burst-source frame or the lab-frame of the engine, where $\sigma_T$ is the Thomson cross section. Actually, $\gamma_{ch}$ corresponds to the minimal Lorentz factor of the truely power-law distributed electrons in the emission region.

\cite{2018ApJ...869..100U} proposed five variation patterns for $\gamma_{ch}$ with radius. However, among these five $\gamma_{ch}$ patterns, there are mainly two polarization evolution patterns corresponding to the hard-to-soft and intensity-tracking spectral peak-energy evolution mode \footnote{The spectral peak energy decays with time all the way for the hard-to-soft mode, while its value is positively correlated with flux for the intensity-tracking mode.}, respectively. For the hard-to-soft mode (denoted as the $i$ model), its $\gamma_{ch}$ reads
\begin{equation}\label{gammai}
\gamma_{ch}(r)=\gamma_{ch}^i(r/r_0)^g,
\end{equation}
where $\gamma_{ch}^i$ is the normalization value of $\gamma_{ch}$ at radius $r_0$. And we take $g=-0.2$ for $i$ model \citep{2018ApJ...869..100U}. For the intensity-tracking mode (denoted as the $m$ model), its $\gamma_{ch}$ reads
\begin{equation}\label{gammam}
\gamma_{ch}(r)=\gamma_{ch}^m\times\begin{cases}
(r/r_m)^g, & \text{$r\leq r_m$}, \\ (r/r_m)^{-g}, & \text{$r\geq r_m$},
\end{cases}
\end{equation}
where $r_m$ is normalization radius of the $m$ model and $\gamma_{ch}^m$ is the value of $\gamma_{ch}$  at $r_m$. And we take $g=1.0$ for $m$ model \citep{2018ApJ...869..100U}.

The equations for the time-resolved and energy-resolved flux density $f_{\nu}$, the Stokes parameters $Q_{\nu}$ and $U_{\nu}$ are shown in Appendix \ref{Calculation equation of polarization}. For the time-integrated ones, the $\overline{f}_{\nu}$, $\overline{Q}_{\nu}$, and $\overline{U}_{\nu}$ are expressed as:
\begin{equation}
\begin{aligned}
&\overline{f}_{\nu}=\frac{\int_{t_5}^{t_{95}}f_{\nu}dt_{obs}}{T_{90}}\\
&\overline{Q}_{\nu}=\frac{\int_{t_5}^{t_{95}}Q_{\nu}dt_{obs}}{T_{90}}\\
&\overline{U}_{\nu}=\frac{\int_{t_5}^{t_{95}}U_{\nu}dt_{obs}}{T_{90}},
\end{aligned}
\end{equation}
where $t_5$ and $t_{95}$ are the times at which the accumulated flux density reaches $5\%$ and $95\%$ of the total flux density, respectively. And the duration of the burst ($T_{90}$) equals to $t_{95}-t_5$. It should be noted that the duration $T_{90}$ of the same burst will vary with the observational energy band, i.e., the lower the observational energy band, the longer the duration \citep{Li_2024}.

Then the time-integrated PD for the models with an aligned field in the radiation region can be expressed as follows.
\begin{equation}
\overline{PD}=\frac{\sqrt{\overline{Q}^2_{\nu}+\overline{U}^2_{\nu}}}{\overline{f}_{\nu}}
\end{equation}
And for the models with a toroidal field, because the Stokes parameter $\overline{U}_{\nu}$ is zero duo to the axial symmetry, its time-integrated PD is
\begin{equation}
\overline{PD}=\frac{\overline{Q}_{\nu}}{\overline{f}_{\nu}}
\end{equation}
The change of the time-resolved and energy-resolved PA ($\Delta$PA) within $T_{90}$ are defined as
\begin{equation}
\Delta PA=PA_{max}-PA_{min}
\end{equation}
where $PA_{max}$ and $PA_{min}$ are the maximum and minimum time-resolved PA values within the $T_{90}$, respectively.

\section{Numerical Results}\label{sec:Numerical Results}

With the model mentioned above, time-integrated spectral polarimetry are investigated and also the influences of the key parameters are studied. Total 7 models are consisered here \citep{2018ApJ...869..100U}. Unless specified, the parameters are taken as their typical or fiducial values: $\Gamma_0=250$, the jet opening angle $\theta_j=0.1$ rad, viewing angle $\theta_V=\theta_j/2=0.05$ rad, $B'_0=30$ G, the starting radius of the emission $r_{on}=10^{14}$ cm, the ending radius of the emission $r_{off}=3\times10^{16}$ cm, $\gamma_{ch}^i=5\times10^4$, $r_0=10^{15}$ cm, $\gamma_{ch}^m=2\times10^5$ and $r_m=2\times10^{15}$ cm \citep{Ghirlanda2018,Lloyd2019,RE2023,2018ApJ...869..100U}. For the aligned magnetic field, its orientation is assumed to be $\delta=\pi/6$ with respect to the projection of the jet axis on the plane of sky. The other parameters for these 7 models are listed in Table. \ref{Model Parameters}.

\begin{deluxetable*}{lcccc}
\tabletypesize{\large}
\tablewidth{20pt}
\tablecaption{Model Parameters.\label{Model Parameters}}
\tablehead{ 
\colhead{Model}
&\colhead{\textcolor{black}{$b$}}
&\colhead{\textcolor{black}{$s$ (MFC)}}
&\colhead{\textcolor{black}{$\gamma_{ch}$ profile}} 
&\colhead{\textcolor{black}{$\alpha_2$}} 
}
\startdata
 $1b_i$ & \textcolor{black}{1.0} &  \textcolor{black}{0 (toroidal)}  & \textcolor{black}{Equation \ref{gammai}} & \textcolor{black}{-1.0}  \\\hline
 $1c_i$ & \textcolor{black}{1.25} &  \textcolor{black}{0 (toroidal)}  & \textcolor{black}{Equation \ref{gammai}} & \textcolor{black}{-1.0}  \\\hline
 $1d_i$ & \textcolor{black}{1.5} &  \textcolor{black}{0 (toroidal)}  & \textcolor{black}{Equation \ref{gammai}} & \textcolor{black}{-1.0}  \\\hline
 $1b_i2$ & \textcolor{black}{1.0} &  \textcolor{black}{0 (toroidal)}  & \textcolor{black}{Equation \ref{gammai}} & \textcolor{black}{-3/2}  \\\hline
 $1b_m$ & \textcolor{black}{1.0} &  \textcolor{black}{0 (toroidal)}  & \textcolor{black}{Equation \ref{gammam}} & \textcolor{black}{-1.0}  \\\hline
 $2b_i$ & \textcolor{black}{1.0} &  \textcolor{black}{0.35 (aligned)}  & \textcolor{black}{Equation \ref{gammai}} & \textcolor{black}{-1.0}  \\\hline
 $2b_m$ & \textcolor{black}{1.0} &  \textcolor{black}{0.35 (aligned)}  & \textcolor{black}{Equation \ref{gammam}} & \textcolor{black}{-1.0}  \\\hline
\enddata
\end{deluxetable*}

\subsection{PD and PA Spectra}

\begin{figure*}
\centering
\includegraphics[scale=0.6]{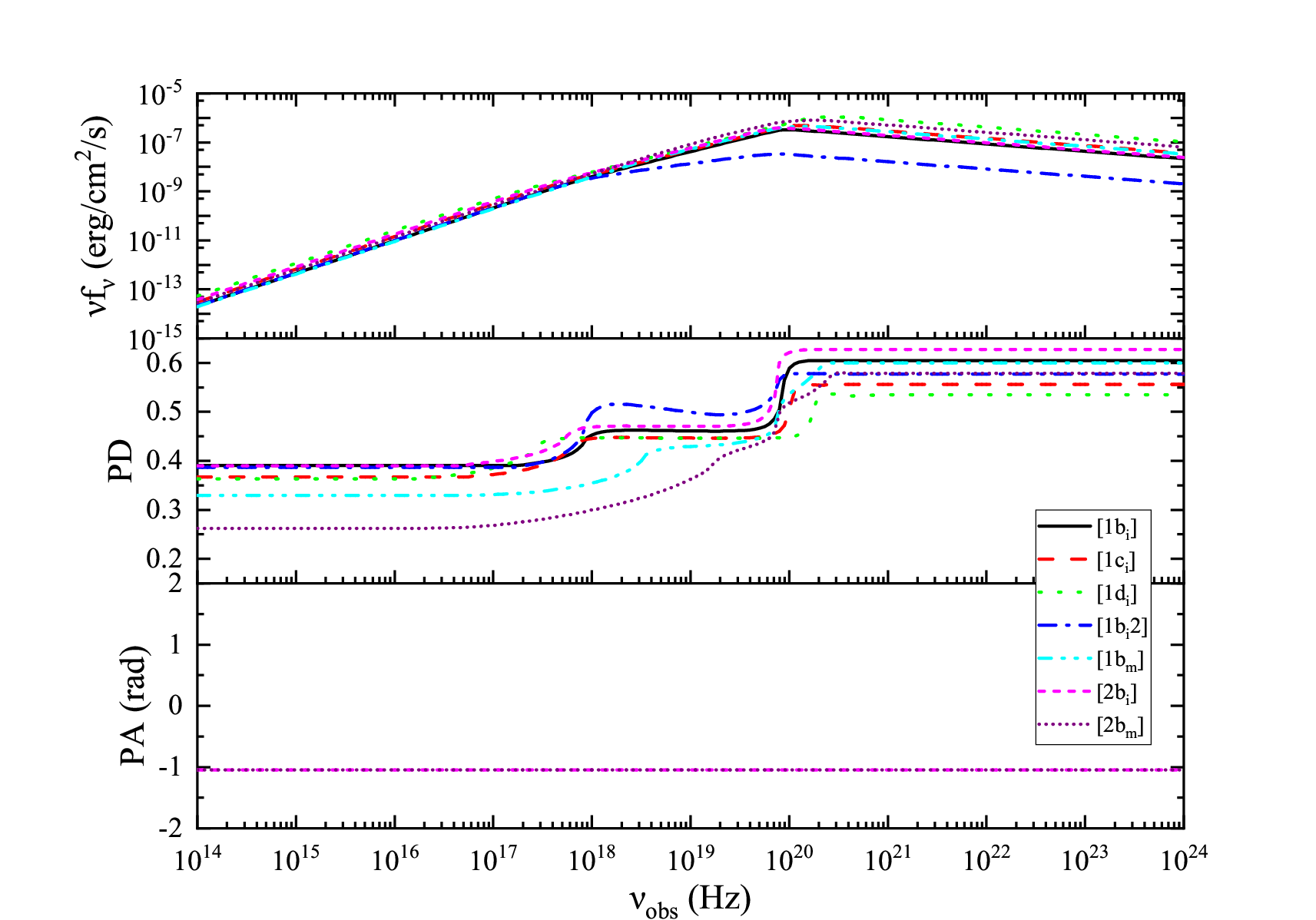}
\caption{Spectra and polarization spectra of the 7 models at the peak time of the corresponding light curve of each model at 300 keV. Top, middle and bottom panels show the spectra, PD spectra and PA spectra, respectively. The black-solid, red-dashed, green-dotted, blue-dash-dotted, cyan-double-dot-dashed, magenta-short-dashed, and purple-short-dashed lines correspond to the models of $[1b_i]$, $[1c_i]$, $[1d_i]$, $[1b_i2]$, $[1b_m]$, $[2b_i]$, and $[2b_m]$, respectively.}
\label{spectrum_tpeak_Uhmmodel}
\end{figure*}

\begin{figure*}
\centering
\includegraphics[scale=0.6]{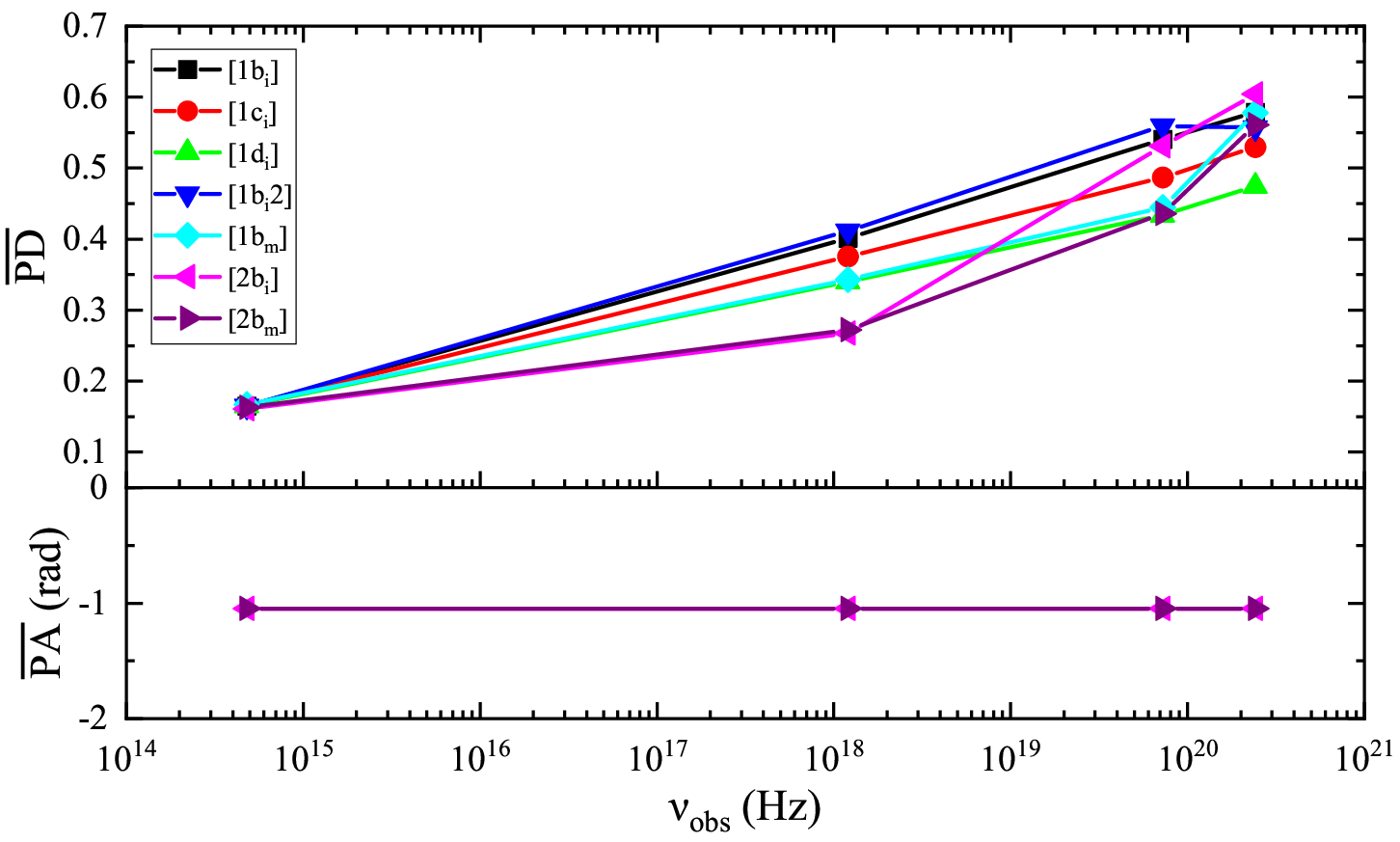}
\caption{Time-integrated polarization spectra for the 7 models. Top and bottom panels show the time-integrated PD and PA spectra, respectively. The black, red, green, blue, cyan, magenta, and purple lines correspond to the models of $[1b_i]$, $[1c_i]$, $[1d_i]$, $[1b_i2]$, $[1b_m]$, $[2b_i]$, and $[2b_m]$, respectively.}
\label{PD_vobs}
\end{figure*}

For the 7 models consisdered, the effects of the magnetic field decaying index $b$, mid-energy photon spectral index $\alpha_2$, $\gamma_{ch}$ patterns, and MFCs ($\Gamma$ variation patterns) on the time-resolved and time-integrated polarization spectra are studied. The time-resolved polarization spectra are investigated at the peak time of the light curve at 300 keV for each model, and the results are shown in Figure. \ref{spectrum_tpeak_Uhmmodel}. Since the change of the spectral indices with frequency would lead to an abrupt increase of the local PD in an ordered magnetic field $\Pi_p$ at $\nu_1$ (i.e., $h\nu_1=E_{break}$) and $\nu_2$ (i.e., $h\nu_2=E_{peak}$), it can be clearly seen that there are three plateaus in the time-resolved PD spectra smoothly connected at the two frequencies. Because the difference between the values of the two critical frequencies $\nu_1$ and $\nu_2$ are small for the $[2b_m]$ model, its second PD plateau is not obvious. The time-integrated polarization spectra are calculated within the $T_{90}$ of each energy band. And the results are shown in Figure. \ref{PD_vobs}. The time-integrated PD generally increases toward the high energy band for all 7 models considered here. Both the time-resolved and time-integrated polarization angles (PAs) are constants with frequency for on-axis observation.

Since the contribution of the local flux density $f_\nu(r)$ at large radii to the total one will be larger for the model with a smaller $b$ and the local PD of the emission from radius $r$ to $r+dr$ ($PD(r)$) would approach its maximum value of $\Pi_p$ at large radii, both the time-resolved and the time-integrated PDs of the $[1b_i]$ model are higher than those of the models $[1c_i]$ and $[1d_i]$. Because the mid-energy photon spectral indices of the $[1b_i]$ and $[1b_i2]$ models are different, leading to a different $\Pi_p$ between the frequencies of $\nu_1$ and $\nu_2$, the time-resolved PDs of the two models show obvious diverse between the two frequencies.

Depending on the evolution of the critical frequency \citep{1998ApJ...497L..17S}, the peak times of the light curves at 300 keV for m model would be larger than that for the i model with the same dynamics. The corresponding radius of the maximum local flux density $f_\nu(r)$ would be larger for the m model. Since the break energy ($E_{break}$) is proportional to $r^3$, the $E_{break}$ would be larger for the m model at the peak times of the light curves at 300 keV. The second PD plateau of the $[1b_m]$ ($[2b_m]$) model begins at a larger frequency compared with that of the $[1b_i]$ ($[2b_i]$) model. Because the time-resolved and time-integrated polarization spectra are similar between the i model and the m model, we use $[1b_i]$ model and $[2b_i]$ model as example to investigate the influences of other key parameters.

\begin{figure*}
\centering
\includegraphics[scale=0.35]{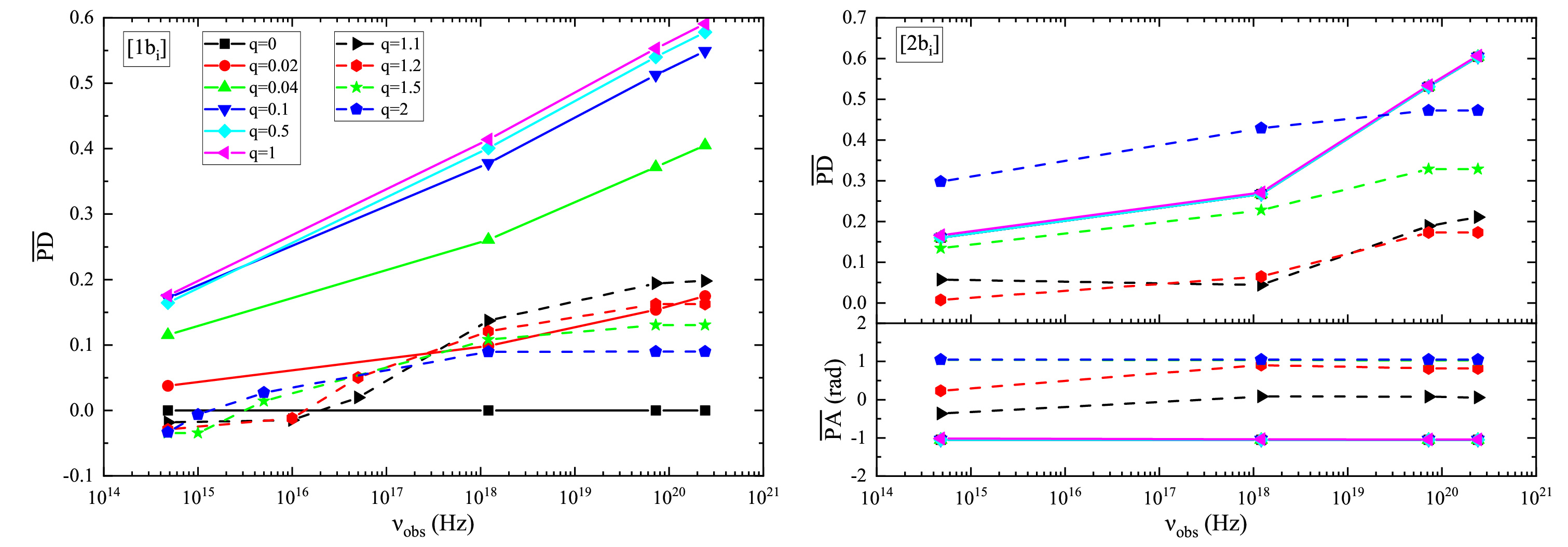}
\caption{Time-integrated polarization spectra with various $q$ values. Left and right panels correspond to the models of $[1b_i]$ and $[2b_i]$, respectively. PA shows abrupt $90\circ$ change when the PD changes its sign for the $[1b_i]$ model, so only the PD spectra are shown in the left panel. In the right panel, top and bottom panels show the time-integrated PD and PA spectra, respectively. The black, red, green, blue, cyan, and magenta solid lines correspond to the on-axis observations with $q=0.00$, 0.02, 0.04, 0.1, 0.5 and 1, respectively. The black, red, green, and blue dashed lines correspond to the off-axis observations with $q=1.1$, 1.2, 1.5 and 2.0, respectively.}
\label{q}
\end{figure*}

Then the effects of the key parameters (i. e., $q$, $\Gamma_0$, and $\theta_j$) on the time-integrated polarization spectra are studied. The time-integrated polarization spectra of the two models with various $q$ are shown in Figure  \ref{q}. For on-axis observations with 
$q\equiv\theta_V/\theta_j\leq1/(\Gamma_0\theta_j)$, 
 the toroidal fields (i.e., corresponding 
to the $[1b_i]$ model) within the 1/Gamma cone are almost closed circles,
 the local polarizations from each closed magnetic field circle cancel with each other, resulting in 
low PDs. Such symmetry is gradually broken with 
the increase of $q$, resulting in the increasing PD. However, for an aligned field (e.g., $[2b_i]$ model), the time-integrated PDs would concentrate around a constant for on-axis observations due to a roughly unchanged asymmetry offerred by the filed. The time-integrated PAs of the $[2b_i]$ model would rotate gradually with energy for slightly off-axis observations (e.g., $q=1.1$ and 1.2), while they show abrupt $90^\circ$ changes slightly beyond $10^{15}$ Hz (i.e., around the Ultraviolet band) for all calculated off-axis observations of the $[1b_i]$ model.

\begin{figure*}
\centering
\includegraphics[scale=0.6]{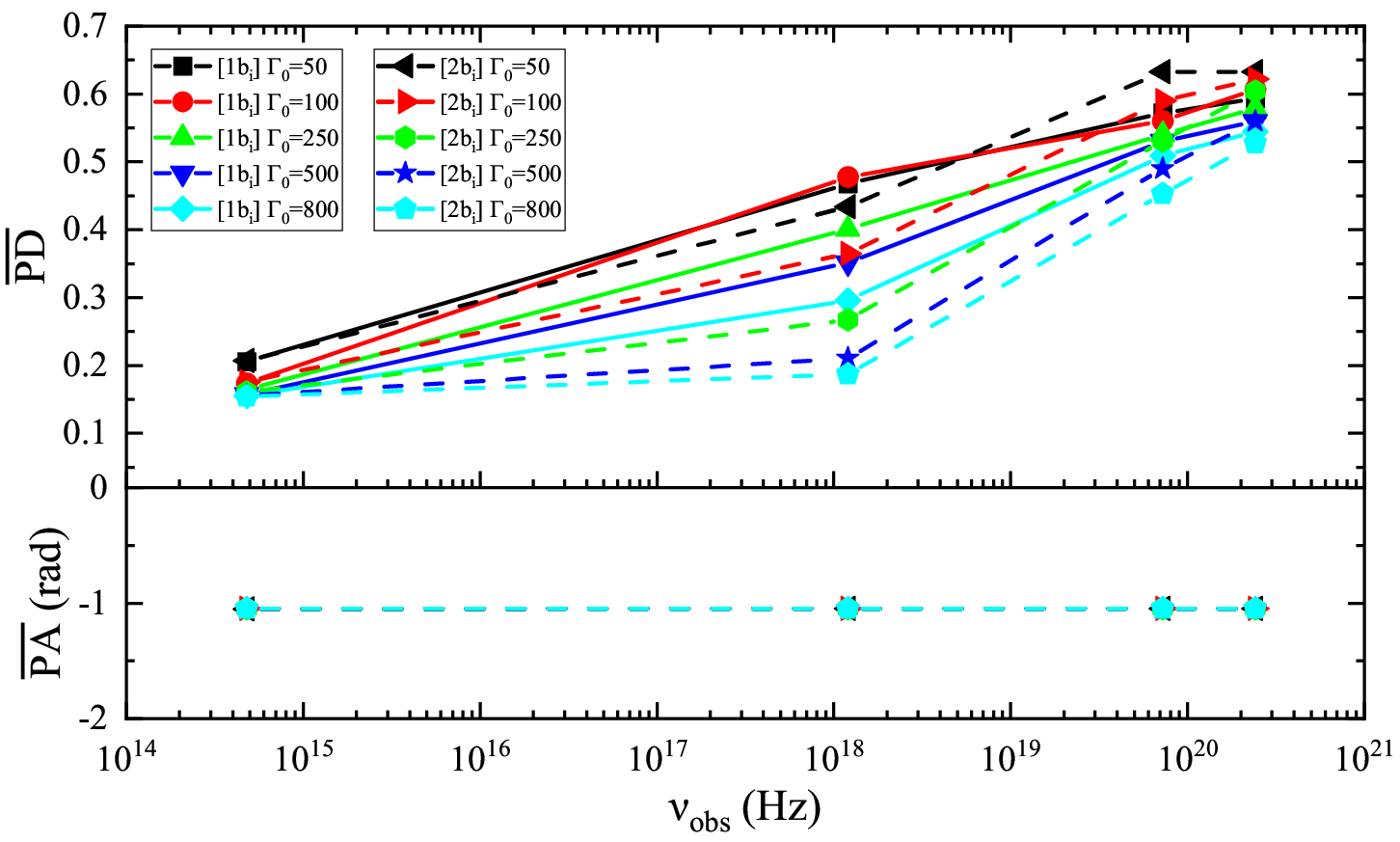}
\caption{Time-integrated polarization spectra with various $\Gamma_0$ values. Top and bottom panels show the time-integrated PD and PA spectra, respectively. The solid and dashed lines correspond to models of $[1b_i]$ and $[2b_i]$, respectively. The black, red, green, blue, and cyan lines correspond to $\Gamma_0=50$, 100, 250, 500, and 800, respectively.}
\label{g0}
\end{figure*}

\begin{figure*}
\centering
\includegraphics[scale=0.6]{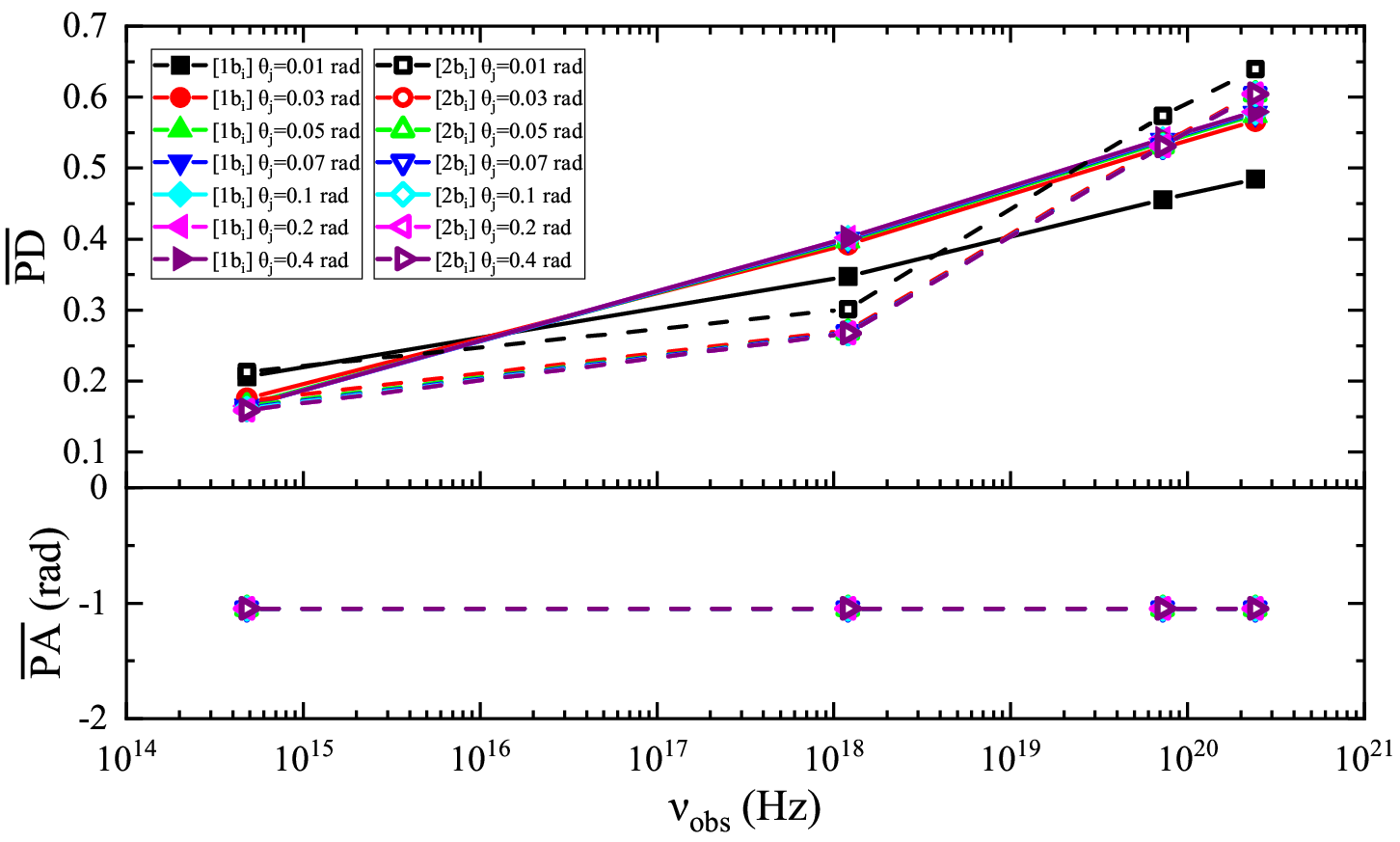}
\caption{Time-integrated polarization spectra with various $\theta_j$ values. Top and bottom panels show the time-integrated PD and PA spectra, respectively. The solid and dashed lines correspond to models of $[1b_i]$ and $[2b_i]$, respectively. The black, red, green, blue, cyan, magenta, and purple lines correspond to $\theta_j=0.01$, 0.03, 0.05, 0.07, 0.1, 0.2, and 0.4 rad, respectively.}
\label{thetaj}
\end{figure*}

The variations of the time-integrated polarization spectra with $\Gamma_0$ are shown in Figure. \ref{g0}. Because with the increase of $\Gamma_0$, the $1/\Gamma$ cone will decrease and the proportion of the radiation within $1/\Gamma$ cone becomes smaller. PD of the emission from inside $1/\Gamma$ cone are higher than that from outside. Therefore, in general the time-integrated PDs will decrease with $\Gamma_0$. The time-integrated PAs remain as a constant with frequency for various $\Gamma_0$ values. The variations of the time-integrated polarization spectra with $\theta_j$ are shown in Figure. \ref{thetaj}. With our calculations, the polarization properties with $\Gamma_0\theta_j\sim1$ would be significantly different from these with $\Gamma_0\theta_j\sim10$, the polarization spectra with $\theta_j=0.01$ rad (corresponding to $\Gamma_0\theta_j=2.5$) show obvious diverse from that of other $\theta_j$ values (corresponding to $\Gamma_0\theta_j\geq7.5$). 

It should be noted that the influence of the parameters on the time-integrated PDs is very tiny in optical band and is most obvious in X-ray band. In optical band, the $T_{95}$ of all 7 models is roughly 10 times of the corresponding end time of the high time-resolved PD phase and the accumlated flux from the end of the high time-resolved PD phase to the $T_{95}$ is about one half of the total flux witin $T_{90}$. Although the PD values at the high time-resolved PD phase for different parameter sets would be different \citep{Li_2024}, the long lasting low polarized phase after the high PD stage would wash out these difference, resulting in a roughly same time-integrated PD ($\sim17\%$) for all 7 models in optical band.

\subsection{$\Delta$PA Spectra}

\begin{figure*}
\centering
\includegraphics[scale=0.4]{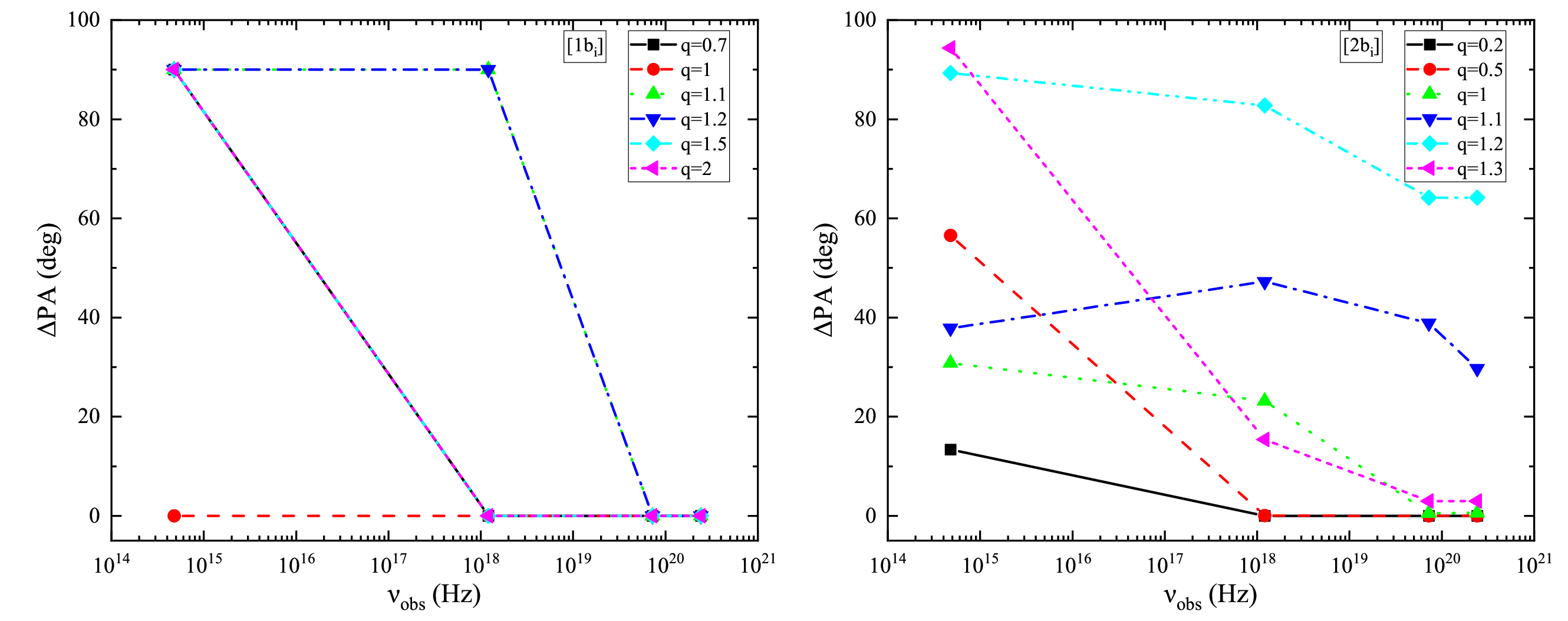}
\caption{PA rotation spectra within $T_{90}$ for various $q$ values. Left and right panels correspond to the models of $[1b_i]$ and $[2b_i]$, respectively. In the left panel, the black, red, green, blue, cyan, and magenta lines correspond to $q=0.7$, 1, 1.1, 1.2, 1.5 and 2, respectively. In the right panel, the black, red, green, blue, cyan, and magenta lines correspond to $q=0.2$, 0.5, 1, 1.1, 1.2 and 1.3, respectively.}
\label{PA_q}
\end{figure*}

The $\Delta$PA spectra within $T_{90}$ are investigated. The $\Delta$PA spectra of the $[1b_i]$ model and of the $[2b_i]$ model for various $q$ values are shown in Figure \ref{PA_q}. In general, PA variation within $T_{90}$ ($\Delta$PA) is larger in a lower energy band for both model. For $[1b_i]$ model with a toroidal field in the radiation region, its PA could stay as a constant or change abruptly by $90^\circ$. In addition to the slightly off-axis observations, the $\Delta$PA could be $90^\circ$ even for on-axis observaton and relatively large off-axis observation in optical band. In X-ray band, the $\Delta$PAs are $90^\circ$ only for slightly off-axis observations (i.e., $q=1.1$ and 1.2). However, the $\Delta$PAs are zero in gamma-ray band with the parameters used here for all the observational angles calculated. For $[2b_i]$ model with an aligned field in the radiation region, the $\Delta$PA could be any value within $0^\circ$ to $90^\circ$. The maximum $\Delta$PA in each energy band would be reached at similar $q$ values. In the optical band, the $\Delta$PAs are nonzero for all the observational angles consisdered, including both on-axis and off-axis observations. The maximum value of $\Delta$PA is $94.36^\circ$ at $q=1.3$. In both X-ray and gamma-ray band, the $\Delta$PA is zero for on-axis observations (except for $q=1$), while it will increase and then decrease with $q$ for the off-axis observation. The maximum value of $\Delta$PA is $82.79^\circ$ at $q=1.2$ in X-rays and is $64.19^\circ$ also at $q=1.2$ in gamma-rays.

\renewcommand\arraystretch{1.0}
\setlength{\tabcolsep}{30pt}
\begin{table}[h]
\large
\caption{The $\triangle$PA spectra of the $[1b_i]$ model at $q=1.2$ for different sets of ($\Gamma_0,\theta_j$) with various fixed product values of $\Gamma_0\theta_j$.}
\label{table 2}
\begin{tabular*}{\hsize}{l|cccc}
\hline
\hline
$\Gamma_0\theta_j=5$ & R Band & 5 keV & 300 keV & 1 MeV  \\\hline
$\Gamma_0=50$ & 90 & 0 & 0 & 0  \\
$\Gamma_0=100$ & 90 & 0 & 0 & 0  \\
$\Gamma_0=250$ & 90 & 90  & 0 & 0  \\\hline
\hline
$\Gamma_0\theta_j=10$ & R Band & 5 keV & 300 keV & 1 MeV  \\\hline
$\Gamma_0=50$ & 90 & 0 & 0 & 0  \\
$\Gamma_0=100$ & 90 & 0 & 0 & 0  \\
$\Gamma_0=200$ & 90 & 90 & 0 & 0  \\
$\Gamma_0=500$ & 90 & 90 & 0 & 0  \\\hline
\hline
$\Gamma_0\theta_j=25$ & R Band & 5 keV & 300 keV & 1 MeV  \\\hline
$\Gamma_0=125$ & 90 & 0 & 0 & 0  \\
$\Gamma_0=250$ & 90 & 90 & 0 & 0  \\
$\Gamma_0=500$ & 90 & 90 & 0 & 0  \\
$\Gamma_0=1000$ & 90 & 90 & 90 & 90  \\\hline
\hline 
$\Gamma_0\theta_j=50$ & R Band & 5 keV & 300 keV & 1 MeV  \\\hline
$\Gamma_0=125$ & 90 & 0 & 0 & 0  \\
$\Gamma_0=250$ & 90 & 0 & 0 & 0  \\
$\Gamma_0=500$ & 90 & 90 & 0 & 0  \\
$\Gamma_0=1000$ & 90 & 90 & 90 & 90  \\\hline
\hline
$\Gamma_0\theta_j=80$ & R Band & 5 keV & 300 keV & 1 MeV  \\\hline
$\Gamma_0=200$ & 90 & 0 & 0 & 0  \\
$\Gamma_0=400$ & 90 & 0 & 0 & 0  \\
$\Gamma_0=800$ & 90 & 90 & 90 & 90  \\
$\Gamma_0=1000$ & 90 & 90 & 90 & 90  \\\hline
\end{tabular*}
\end{table}

\renewcommand\arraystretch{1.0}
\setlength{\tabcolsep}{30pt}
\begin{table}[h]
\large
\caption{The $\triangle$PA spectra of the $[2b_i]$ model at $q=1.2$ for different sets of ($\Gamma_0,\theta_j$) with various fixed product values of $\Gamma_0\theta_j$.}
\label{table 3}
\begin{tabular*}{\hsize}{l|cccc}
\hline
\hline
$\Gamma_0\theta_j=5$ & R Band & 5 keV & 300 keV & 1 MeV  \\\hline
$\Gamma_0=50$ & 20.29 & 18.29 & 27.86 & 27.86  \\
$\Gamma_0=100$ & 46.29 & 19.1 & 26.15 & 28.04  \\
$\Gamma_0=250$ & 61.31 & 21.5  & 25.85 & 29.06  \\\hline
\hline
$\Gamma_0\theta_j=10$ & R Band & 5 keV & 300 keV & 1 MeV  \\\hline
$\Gamma_0=50$ & 35.45 & 41.91 & 14.19 & 14.19  \\
$\Gamma_0=100$ & 63.82 & 46.47 & 14.02 & 14.02  \\
$\Gamma_0=200$ & 82.23 & 45.63 & 17.36 & 14.02  \\
$\Gamma_0=500$ & 84.33 & 55.3 & 25.04 & 14.09  \\\hline
\hline
$\Gamma_0\theta_j=25$ & R Band & 5 keV & 300 keV & 1 MeV  \\\hline
$\Gamma_0=125$ & 169.33$^\ast$ & 81.82 & 64.76 & 64.76  \\
$\Gamma_0=250$ & 89.33 & 82.79 & 64.19 & 64.19  \\
$\Gamma_0=500$ & 89.22 & 81.43 & 69.67 & 64.95  \\
$\Gamma_0=1000$ & 89.21 & 83.18 & 73.14 & 66.01  \\\hline
\hline 
$\Gamma_0\theta_j=50$ & R Band & 5 keV & 300 keV & 1 MeV  \\\hline
$\Gamma_0=125$ & 168.75$^\ast$ & 87.83 & 83.24 & 83.24  \\
$\Gamma_0=250$ & 176.83$^\ast$ & 87.77 & 79.15 & 79.15  \\
$\Gamma_0=500$ & 178.33$^\ast$ & 87.96 & 78.09 & 78.09  \\
$\Gamma_0=1000$ & 89.87 & 87.04 & 80.02 & 80.02  \\\hline
\hline
$\Gamma_0\theta_j=80$ & R Band & 5 keV & 300 keV & 1 MeV  \\\hline
$\Gamma_0=200$ & 89.07 & 90.31 & 89.36 & 89.36  \\
$\Gamma_0=400$ & 178.39$^\ast$ & 89.17 & 85.57 & 85.57  \\
$\Gamma_0=800$ & 178.04$^\ast$ & 89.06 & 84.39 & 84.39  \\
$\Gamma_0=1000$ & 163.35$^\ast$ & 88.79 & 85.2 & 85.2  \\\hline
\end{tabular*}
\tablecomments{The superscript $^\ast$ indicates that there are three $\sim90^\circ$ PA rotations within $T_{90}$.}
\end{table}

Because in general the PA rotation for $q=1.2$ is most significant in each energy band, we take $q=1.2$ in the following to study the effects of the $\Gamma_0$ and $\theta_j$ on the $\Delta$PA spectra. We first check whether or not $\Delta$PA is only affected by the product $\Gamma_0\theta_j$ and does not depend on the specific values of $\Gamma_0$ and $\theta_j$. The $\Delta$PA values for different sets of ($\Gamma_0,\theta_j$) are listed in Table \ref{table 2} for $[1b_i]$ model and Table \ref{table 3} for $[2b_i]$ model. For $[1b_i]$ model with a fixed product value of $\Gamma_0\theta_j$, the energy band with abrupt $90^\circ$ PA rotation extends from single optical band to more energy bands with the increase of $\gamma_0$. So for $[1b_i]$ model with a toroidal magnetic field in its radiation region, the $\Delta$PA would depend on the concrete values of both $\Gamma_0$ and $\theta_j$ in the energy bands above X-rays, and for all the parameter sets considered here the $\Delta$PA is $90^\circ$ in optical band. For $[2b_i]$ model with an aligned field in the emission region, $\triangle$PA values in the optical band depend on the specific values of $\Gamma_0$ and $\theta_j$. However in the X-ray and gamma-ray bands, the difference of $\triangle$PA for the various sets of ($\Gamma_0,\theta_j$) with a fixed product value of $\Gamma_0\theta_j$ is in general less than $10^\circ$. Therefore, the $\triangle$PA values in the X-ray and gamma-ray bands are mainly dependent on the product value of $\Gamma_0\theta_j$ and independent on the concrete values of both $\Gamma_0$ and $\theta_j$ for $[2b_i]$ model.

\begin{figure*}
\centering
\includegraphics[scale=0.4]{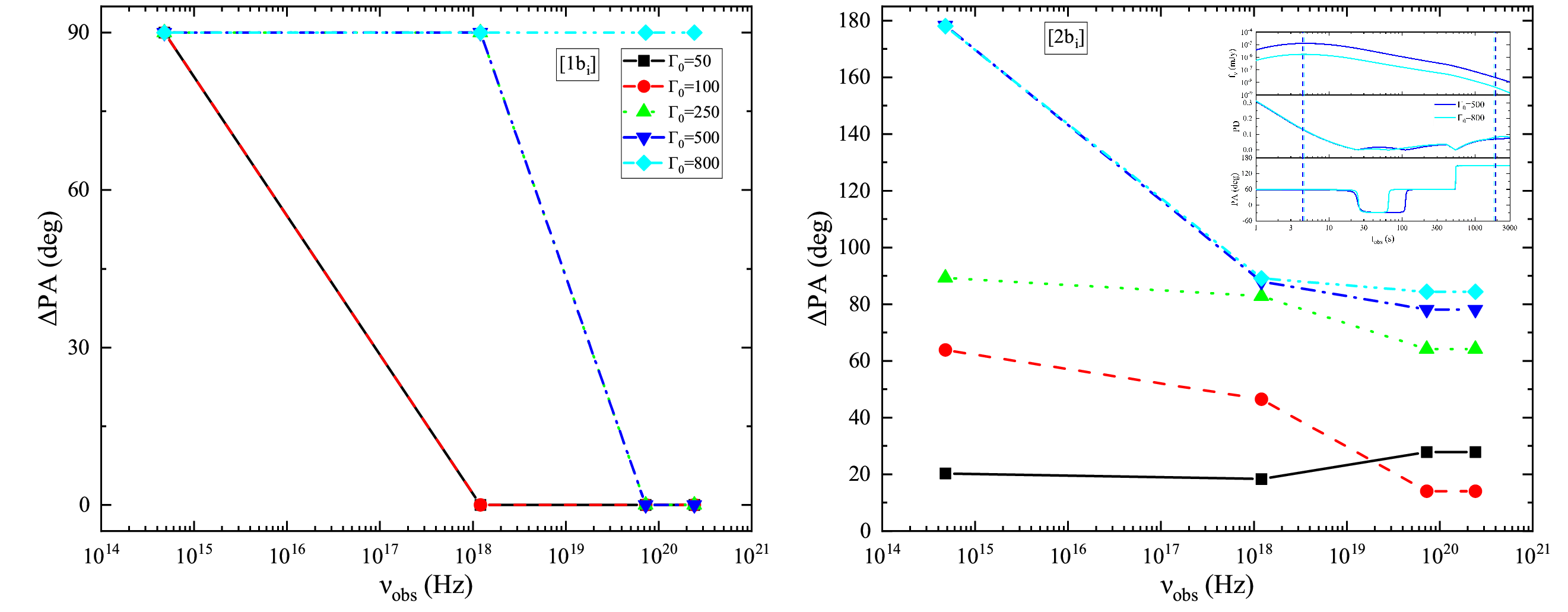}
\caption{PA rotation spectra of $q=1.2$ for various $\Gamma_0$ values. The left and right panels correspond to the models of $[1b_i]$ and $[2b_i]$, respectively. The black, red, green, blue, and cyan lines correspond to $\Gamma_0=50$, 100, 250, 500 and 800, respectively. For the $[2b_i]$ model, the $\Delta$PAs of the $\Gamma_0=500$ and 800 in the optical band are close to $180^\circ$, that is because there are three $\sim90^\circ$ PA rotations within $T_{90}$. In the upper right corner of the right panel, the corresponding light curves and polarization curves with three PA rotations within $T_{90}$ for  $\Gamma_0=500$ and 800 are shown, and reference lines, with the same color as its light curves, indicate the $T_{90}$ ranges in optical band.}
\label{PA_g0}
\end{figure*}

With the above conclusions, the dependences of $\triangle$PA spectra on both $\Gamma_0$ and $\theta_j$ are considered. The $\Delta$PA spectra with varing $\Gamma_0$ values are shown in Figure \ref{PA_g0}. For the $[1b_i]$ model, with the increase of the $\Gamma_0$, the energy band with abrupt $90^\circ$ PA rotation extends from only optical band to all four energy band. For the $[2b_i]$ model, in general, $\Delta$PA will be larger for a larger $\Gamma_0$ value in each energy band. The $\Delta$PA would decrease with observational energy for most of the $\Gamma_0$ values considered, except $\Gamma_0=50$. It is worth noting that the $\Delta$PAs for  $\Gamma_0=500$ and 800 are close to $180^\circ$ in the optical band, which is due to the existence of three $\sim90^\circ$ rotations of PA within $T_{90}$. The corresponding PA curves for the two cases are shown in the inset figure of Figure \ref{PA_g0}.

\begin{figure*}
\centering
\includegraphics[scale=0.4]{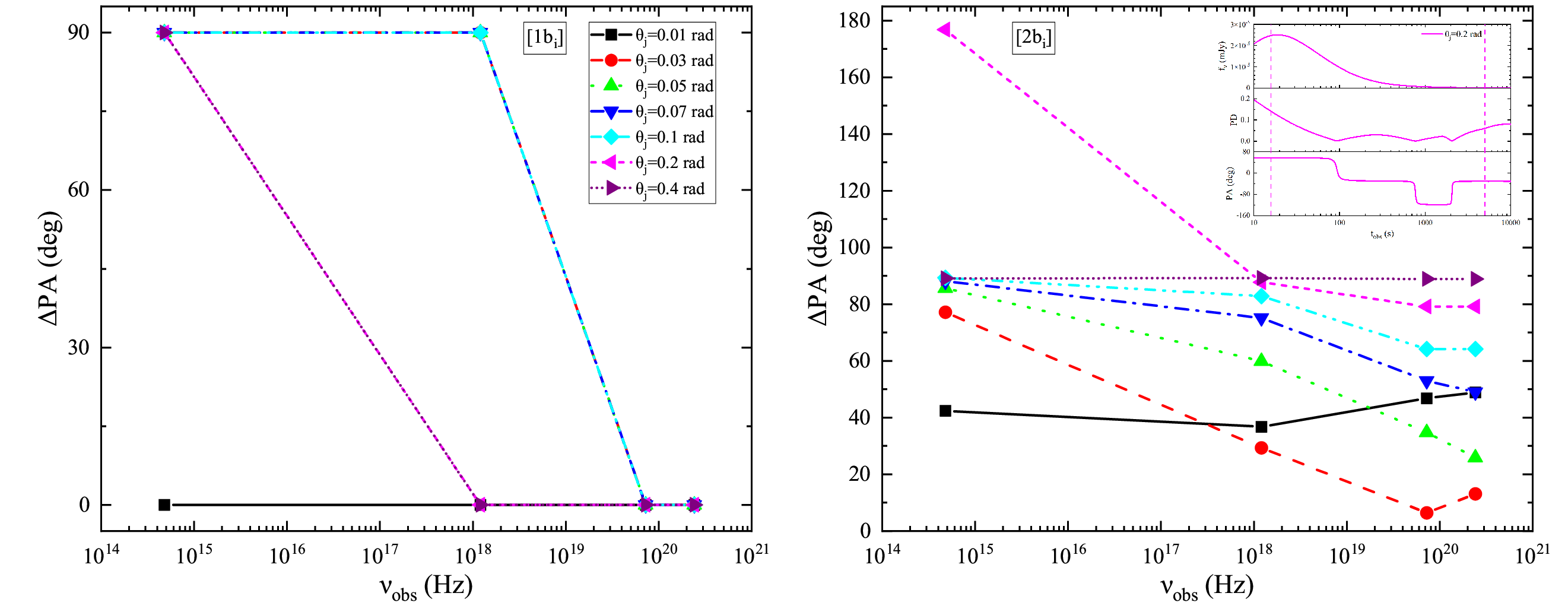}
\caption{PA rotation spectra of $q=1.2$ for various $\theta_j$ values. The left and right panels correspond to the models of $[1b_i]$ and $[2b_i]$, respectively. The black, red, green, blue, cyan, magenta, and purple lines correspond to $\theta_j=0.01$, 0.03, 0.05, 0.07, 0.1, 0.2, and 0.4 rad, respectively. For the $[2b_i]$ model, the $\Delta$PA of the $\theta_j=0.2$ rad in the optical band is close to $180^\circ$, it is also because there are three $90^\circ$ PA rotations within $T_{90}$. In the upper right corner of the right panel, light curves and polarization curves with three PA rotations within $T_{90}$ for $\theta_j=0.2$ rad are shown, and reference lines indicate the $T_{90}$ range in optical band.}
\label{PA_thetaj}
\end{figure*}

The $\Delta$PA spectra with varing $\theta_j$ values are shown in Figure \ref{PA_thetaj}. For the $[1b_i]$ model, the PAs in the gamma-ray band remain constant within $T_{90}$. Below the gamma-ray band, the lower the energy band, the larger the range of $\theta_j$ with $\Delta$PA$=90^\circ$. For the $[2b_i]$ model, the $\Delta$PA will increase with $\theta_j$ in general. For each $\theta_j$ value, the $\Delta$PA will decrease with the increase of the observational energy band except for the narrow jets with $\theta_j=0.01$ rad and 0.03 rad. As above in Figure \ref{PA_g0}, there are also three PA rotations within $T_{90}$ for $\theta_j=0.2$ rad and the $\Delta$PA is roughly $180^\circ$.

\section{Conclusions and Discussion}\label{sec:Conclusions and Discussion}

The polarization spectra down to the optical band of the photosphere model had been investigated in \cite{Parsotan_2022}. However, it is not considered in the magnetic reconnection model so far. In this paper, under frame work of the magnetic reconnection model both the time-resolved and the time-integrated polarization spectra (including the PD spectra, the PA spectra and the $\Delta$PA spectra) are investigated. Total 7 models are considered and the influences of the key parameters on polarization spectra are studied. 

In general, there are three PD plateaux in the time-resolved PD spectra, smoothly connected at the break energy and the peak energy. For on-axis observations, the time-integrated PD would in general increase with frequency and the result is independent of the key parameters, such as jet opening angle, the viewing angle and the bulk Lorentz factor. The time-integrated PDs would concentrate around $17\%$ in optical R-band here and the influence of the parameters are very tiny. With respect to the different values of the parameters, the ranges of the time-integrated PDs would be ($20\%$, $35\%$) in X-ray, ($45\%$, $55\%$) in 300 keV, and ($47\%$, $60\%$) in 1 MeV, respectively. However, the variation trend of the time-integrated PD with frequency is not monotonic in the photosphere model \citep{Parsotan_2022}. The first main difference found here between the two models would be the variation trend of the time-integrated PD with frequency, which provides a new method to distinguish between the two models.

For on-axis observations, the time-integrated PA would stay as a constant with frequency for the magnetic reconnection model, while it rotates randomly for the photosphere model \citep{Parsotan_2022}. For the magnetic reconnection model with off-axis observations, the time-integrated PD would decrease to zero and then increase with frequency for the toroidal field in the radiation region and the time-integrated PA would change abruptly by $90^\circ$ when the time-integrated PD reaches zero. Such abrupt PA rotations would happen roughly in the UltraViolet energy band. For the aligned field case, independent of the observational geometry the time-integrated PDs would in general increase with frequency, and the time-integrated PA would rotate gradually with frequency only for slightly off-axis observations. Such gradual PA rotation would happen from optical band to X-rays. Therefore, in general the variation trend of the time-integrated PA with frequency is not random for the magnetic reconnection model, while it varies randomly for the photosphere model \citep{Parsotan_2022}. This is the second difference found here between the magnetic reconnection model and the photophere model.

The PA rotation spectra are also studied. In general, the rotation value of PA within $T_{90}$ will decrease with the increase of the observational energy band. Most significant PA rotation would happen for slightly off-axis observations in each energy band, which is consistent with our former study in gamma-ray band  \citep{WL_2023a}. For on-axis observations, PA rotations will only happen in optical band. For the models with a toroidal field in its radiation region, the change of the PA can only be abrupt $90^\circ$ and most of the abrupt $90^\circ$ PA rotations within the burst duration happen in the energy bands below X-rays. Such abrupt PA rotations are quite rare in gamma-ray band and there is only one parameter set under which there is one abrupt $90^\circ$ PA rotation in gamma-ray band during the burst. For the models with an aligned field in the emission region, the rotation value of PA can range from $0^\circ$ to $90^\circ$ and there can be more than once $90^\circ$ PA rotations within the burst duration. With the increase of the $\Gamma_0$ (or $\theta_j$), the rotation value of the time-resolved  PA will roughly increase. Even more, there are three $\sim90^\circ$ PA rotaions within the burst duration in optical band for some parameter sets. In both X-ray and gamma-ray bands, $\triangle$PA values with an aligned field in the radiation region are determined by the product value of $\Gamma_0\theta_j$ and is independent of the concrete values of both $\Gamma_0$ and $\theta_j$. However, for the model with a toroidal field in its emission region, the results depend on both values of the $\Gamma_0$ and $\theta_j$, so does for the optical band in the model with an aligned field. In our study, the jet is assumed to be uniform and axisymmetric. It should be noted that the gradual PA rotation would happen only for the cases with an aligned field in the radiation region for off-axis observations in our model. However, PA of the on-axis observation could also change continuously for non-axisymmetric patchy jets \citep{2024MNRAS.52712178G}.

\begin{acknowledgements}
We thank the anonymous referee for useful suggestions. This work is supported by the National Natural Science Foundation of China (grant Nos. 12473040, and 11903014). M.X.L also would like to appreciate the financial support from Jilin University. 
\end{acknowledgements}

\appendix
\restartappendixnumbering

\section{The time-resolved and energy-resolved Stokes parameters}\label{Calculation equation of polarization}

According to \cite{Lan_2020,SL2024}, the flux density $f_{\nu}$, the time-resolved and energy-resolved Stokes parameters $Q_{\nu}$ and $U_{\nu}$ will be expressed as:
\begin{equation}\label{fQU}
\begin{aligned}
&f_{\nu}=\frac{1+z}{4\pi D_L^2}\int\frac{\mathcal{D}^2}{\Gamma}\frac{c}{4\pi r}NP'_0H_{en}(\nu')\sin\theta'_Bd\phi dt\\
&Q_{\nu}=\frac{1+z}{4\pi D_L^2}\int\frac{\mathcal{D}^2}{\Gamma}\frac{c}{4\pi r}NP'_0H_{en}(\nu')\sin\theta'_B\Pi_p\cos2\chi_pd\phi dt\\
&U_{\nu}=\frac{1+z}{4\pi D_L^2}\int\frac{\mathcal{D}^2}{\Gamma}\frac{c}{4\pi r}NP'_0H_{en}(\nu')\sin\theta'_B\Pi_p\sin2\chi_pd\phi dt,
\end{aligned}
\end{equation}
where $D_L$ is the luminosity distance of the source, $N=\int R_{inj}dt/\Gamma$ is the total electron number in the shell,  and $P'_0=3\sqrt{3}m_ec^2\sigma_TB'/(32q_e)$ represents the magnitude of the spectrum \citep{1979rpa..book.....R}. The $R_{inj}$ is the electron injection rate. And $\phi$ is the angle in the plane of sky between the projection of the jet axis and projection of the radial direction of a local fluid element. The $\Pi_p$ is the local PD and reads:
\begin{equation}\label{localPD}
\Pi_p=\tilde{\alpha}/(\tilde{\alpha}-2/3),
\end{equation}
where $\tilde{\alpha}$ is the photon spectral index. $\tilde{\alpha}=\alpha_1$ for $\nu'< \nu'_1$, $\tilde{\alpha}=\alpha_2$ for $\nu'_1< \nu'< \nu'_2$, and $\tilde{\alpha}=\beta$ for $\nu'> \nu'_2$. The $\chi_p$ is the local PA and its expression can be found in \cite{2016ApJ...816...73L}.

For an aligned field, the time-resolved and energy-resolved PD ($PD$) and PA ($PA$) read:
\begin{equation}
PD=\frac{\sqrt{Q^2_\nu+U^2_\nu}}{f_\nu},
\end{equation}
and
\begin{equation}
PA_{pre}=\frac{1}{2}\arctan\left(\frac{U_\nu}{Q_\nu}\right).
\end{equation}
The final PA ($PA$) needs to be adjusted according to the value of the Stokes parameter. When $Q_{\nu}>0$, the final PA equals to $PA_{pre}$; when $Q_{\nu}<0$, if $U_{\nu}>0$ then the PA is $PA=PA_{pre}+\pi/2$, if $U_{\nu}<0$ then the PA is $PA=PA_{pre}-\pi/2$ \citep{2018ApJ...860...44L}.

For a toroidal field, $U_{\nu}$ is always 0 because of axial symmetry. So the time-resolved PD ($\Pi$) reads:
\begin{equation}
PD=\frac{Q_\nu}{f_\nu}.
\end{equation}
PA will rotate $\pi/2$ when PD changes its sign.

Considering the equal arrival time surface (EATS) effect, the observer time $t_{obs}$ is related to the burst-source time $t$ as follows \citep{1998ApJ...494L..49S,2016ApJ...825...97U}:
\begin{equation}
t_{obs}=\left[t-\frac{r}{c}\cos\theta-t_{on}+\frac{r_{on}}{c}\right](1+z),
\end{equation}
The shell begins to radiate photons at radius $r_{on}$ at a burst-source time $t_{on}$.

\section{Spectral polarimetry at the peak time of the light curve at 300 keV}\label{SPatpeaktime}

The impact of the bulk Lorentz factor $\Gamma_0$, jet opening angle $\theta_j$, viewing angle $\theta_V$, and MFCs on the spectra and polarization spectra at the peak time of the light curve in each model at 300 keV are investigated. The model parameters are shown in Table. \ref{Model Parameters}. The orientation of the aligned magnetic field is set to be $\delta=\pi/6$. The other parameters are fixed as follows: $B_0=30$ G, $r_{on}=10^{14}$ cm, $r_{off}=3\times10^{16}$ cm, $\gamma_{ch}^i=5\times10^4$, and $r_0=10^{15}$ cm \citep{2018ApJ...869..100U}.

\begin{figure*}
\centering
\includegraphics[scale=0.6]{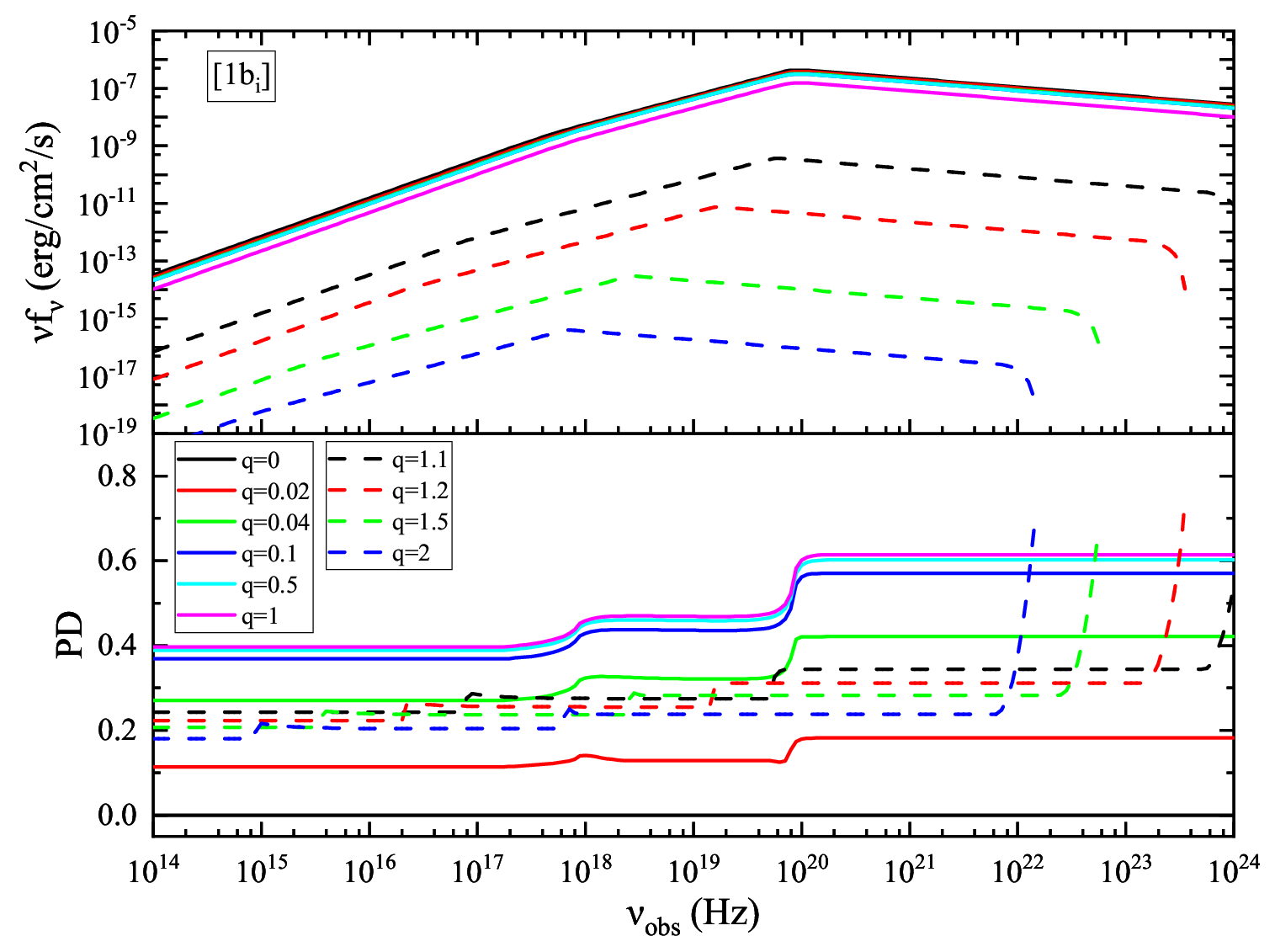}
\caption{Spectra and polarization spectra of the $[1b_i]$ model at the peak time of the light curve at 300 keV for various $q$ values. Top and bottom panels show the spectra and PD spectra, respectively. The black, red, green, blue, cyan, and magenta solid lines correspond to $q=0$, 0.02, 0.04, 0.1, 0.5 and 1, respectively. The black, red, green, and blue dashed lines correspond to $q=1.1$, 1.2, 1.5 and 2.0, respectively.}
\label{spectrum_q_toroidal}
\end{figure*}

\begin{figure*}
\centering
\includegraphics[scale=0.6]{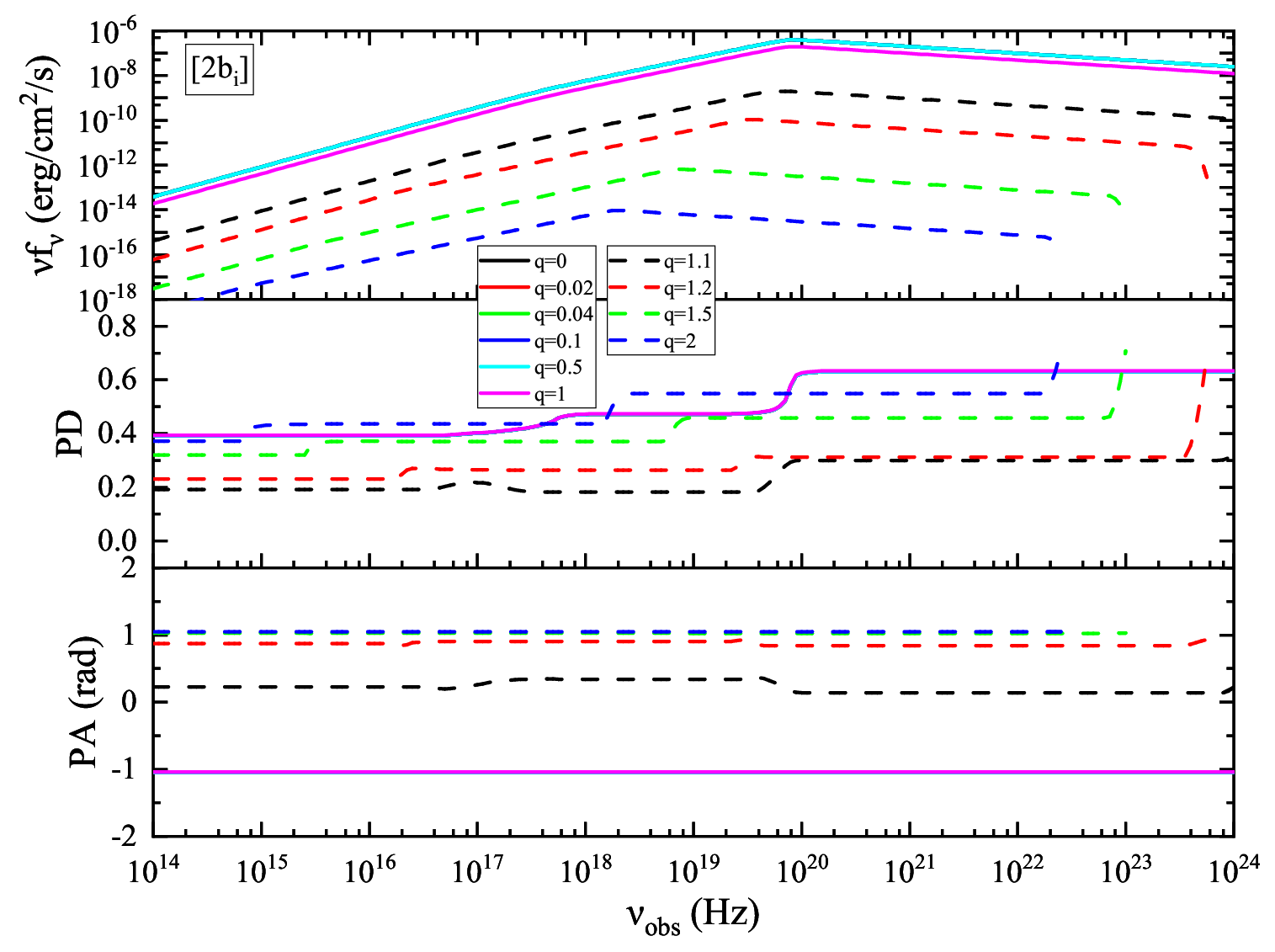}
\caption{Spectra and polarization spectra of the $[2b_i]$ model at the peak time of the light curve at 300 keV for various $q$ values. Top, middle and bottom panels show the spectra, PD spectra and PA spectra, respectively. The black, red, green, blue, cyan, and magenta solid lines correspond to $q=0$, 0.02, 0.04, 0.1, 0.5 and 1, respectively. The black, red, green, and blue dashed lines correspond to $q=1.1$, 1.2, 1.5 and 2.0, respectively.}
\label{spectrum_q_aligned}
\end{figure*}

We take $\Gamma_0=250$ and $\theta_j=0.1$ rad, then the numerical results with various $q$ values are shown in Figures \ref{spectrum_q_toroidal} and \ref{spectrum_q_aligned}. Similar to the time-integrated polarization spectra, with the increase of $q$, time-resolved PDs of a toroidal field will increase when $q\leq1/(\Gamma_0\theta_j)$ and decrease when $q>1$. The spectra and polarization spectra of an aligned field for on-axis observations are very similar. However, for the off-axis observations, time-resolved PDs will decrease and then increase with $q$, while time-resolved PAs changes gradually with frequency for slightly off-axis observations (i.e., $q=1.1$, 1.2).

For off-axis observations (i.e., $q=1.1$, 1.2, 1.5, and 2.0), the flux density disappears suddenly and correspondingly the time-resolved PDs would rise with frequency steeply. This is because the observed frequency $\nu_{obs}$ exceeds the maximum frequency $\nu_{max}=(3\mathcal{D}q^2)/[(1+z)\sigma_Tm_ec]$ of the synchrotron emission. Taking the $[1b_i]$ model with a toroidal magnetic field in its radiation region as an example, because the bulk Lorentz factor $\Gamma$ is constant, the Doppler factor $\mathcal{D}$ would increase with the radius $r$. Thus, on an EATS, $\nu_{max}$ at small radius is smaller, so the flux density vanishes at lower frequencies from small radius and the flux density would be from the large radius at higher frequencies. The shrinking radiation region with frequency would lead to the steep increase of the time-resolved PD with frequency at high energy band. The cease frequency of the flux (corresponding to the maximum frequecy) for a larger $q$ case is smaller. That is because the peak time of the light curve at 300 keV with a larger $q$ value is larger, the values of $\theta$ on the corresponding EATS become relatively larger, leading to a smaller Doppler factor, hence leading to a smaller maximum frequency. The peak time of the light curve for a larger observational angle would be larger, leading to a smaller Doppler factor in the radiation region and to a smaller maximum frequency. So the flux density disappears at a smaller frequency for a larger observational angle.

\begin{figure*}
\centering
\includegraphics[scale=0.6]{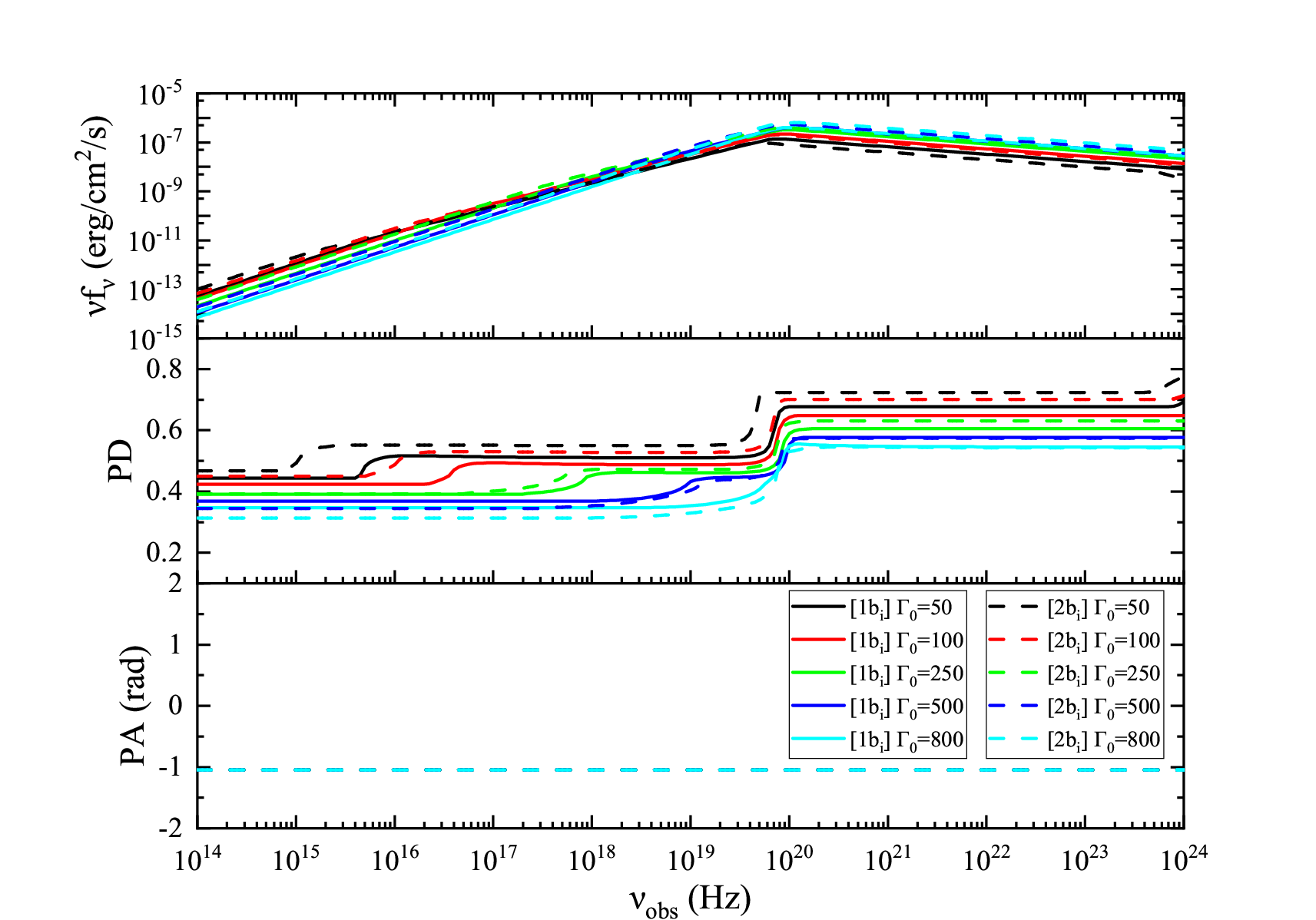}
\caption{Spectra and polarization spectra at the peak time of the corresponding light curve at 300 keV for various bulk Lorentz factor $\Gamma_0$. Top, middle and bottom panels show the spectra, PD spectra and PA spectra, respectively. The solid and dashed lines correspond to models of $[1b_i]$ and $[2b_i]$, respectively. The black, red, green, blue, and cyan lines correspond to $\Gamma_0=50$, 100, 250, 500, and 800, respectively.}
\label{spectrum_tpeak_g0}
\end{figure*}

\begin{figure*}
\centering
\includegraphics[scale=0.6]{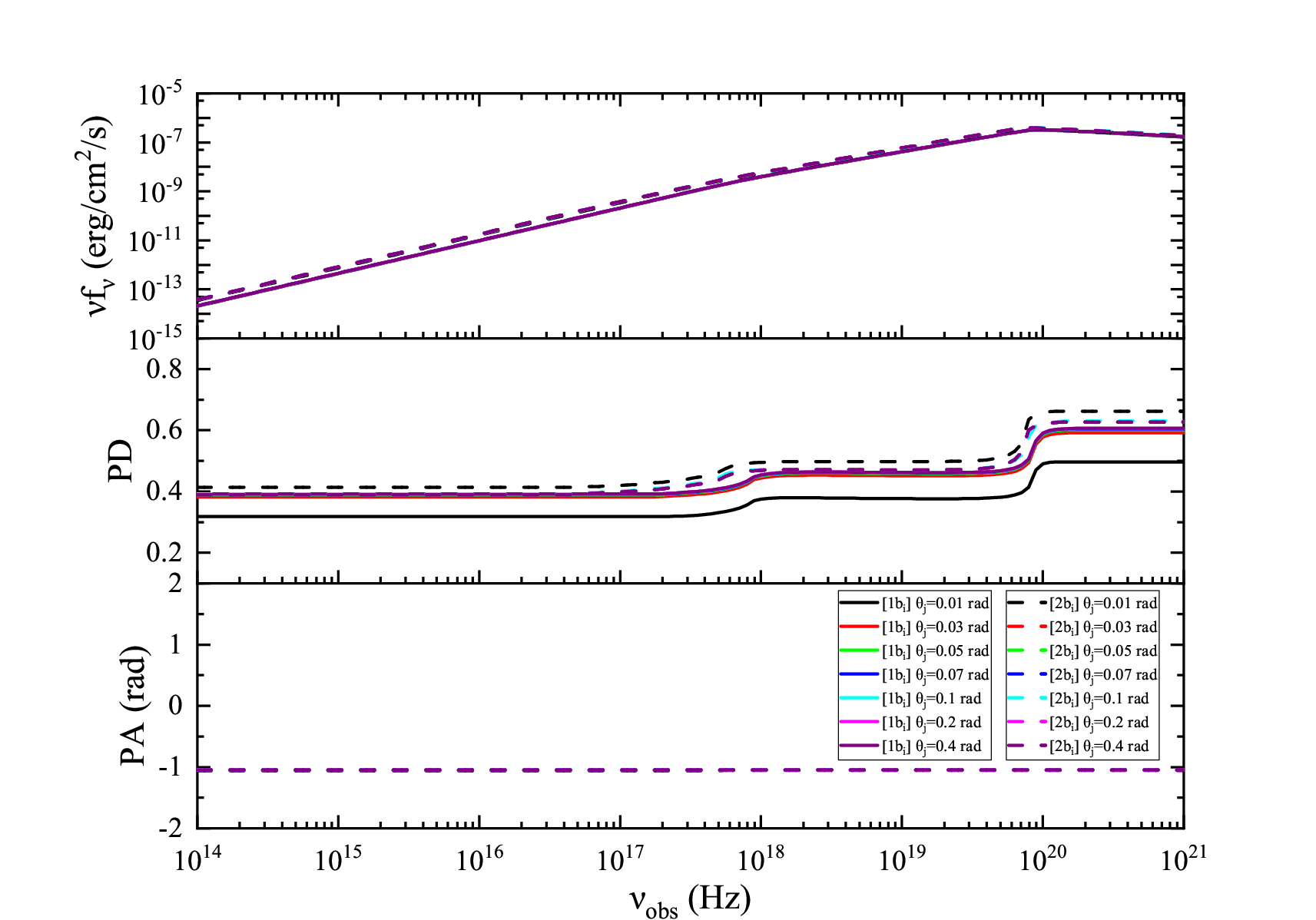}
\caption{Spectra and polarization spectra at the peak time of the corresponding light curve at 300 keV for various jet opening angle $\theta_j$. Top, middle and bottom panels show the spectra, PD spectra and PA spectra, respectively. The solid and dashed lines correspond to models of $[1b_i]$ and $[2b_i]$, respectively. The black, red, green, blue, cyan, magenta, and purple lines correspond to $\theta_j=0.01$, 0.03, 0.05, 0.07, 0.1, 0.2, and 0.4 rad, respectively.}
\label{spectrum_tpeak_thetaj}
\end{figure*}

We take $\theta_j=0.1$ rad and $q=0.5$, and the numerical results with various $\Gamma_0$ values are shown in Figure. \ref{spectrum_tpeak_g0}. At the peak time of the light curve at 300 keV, the cooling is fast, so $\nu_{cool}$ corresponds to $E_{break}$. Due to $\nu_{cool}\propto\gamma_{cool}^2\propto\Gamma^4$, $E_{break}$ would move towards the higher energy band with the increase of $\Gamma_0$, leading to a shorter second PD plateau with $\Gamma_0$. Because the proportion of the radiation within $1/\Gamma$ cone of high local PD would decrease with $\Gamma_0$, time-resolved PD will decrease with $\Gamma_0$. Finally, we take $\Gamma_0=250$ rad and $q=0.5$, and the numerical results for various $\theta_j$ values are shown in Figure. \ref{spectrum_tpeak_thetaj}. The time-resolved PD values of the $[1b_i]$ model are similar for all calculated $\theta_j$ values except $\theta_j=0.01$ rad. Since the syntropy of the toroidal magnetic field within the region of $1/\Gamma$ cone is lower when $q\leq1/(\Gamma_0\theta_j)$, for $\theta_j=0.01$ rad and $q=0.5\sim1/(\Gamma_0\theta_j)$ of the $[1b_i]$ model its time-resolved PDs are relatively lower than other $\theta_j$ values.

\listofchanges

\bibliography{ms_arXiv}

\end{document}